\documentclass[aps,prc,print,showpacs,superscriptaddress,onecolumn,
amsmath,amssymb,longbibliography]{revtex4}
\usepackage{amsfonts}
\usepackage{txfonts}
\usepackage{amsmath}
\usepackage{float}
\usepackage[colorlinks,breaklinks,linkcolor=blue,anchorcolor=blue,citecolor=blue,urlcolor=blue,
dvipdfm
]{hyperref}

\usepackage{graphicx}
\usepackage{dcolumn}
\usepackage{bm}
\usepackage{amssymb}
\usepackage{mathrsfs}
\usepackage[sort&compress]{natbib}
\usepackage{bm,hyperref}
\usepackage{subfigure}
\usepackage{amsmath}

\begin{document}

\title{Hydrodynamical response of plane correlation in Pb+Pb collisions at $\sqrt{s_\text{NN}}$=2.76 TeV}

\author{Jing Yang}
\email[]{yangjdut@gmail.com}
\affiliation{School of Physics, Changchun Normal University, Changchun, Jilin 130032, China}

\author{Yong Zhang}
\email[]{zhy913@jsut.edu.cn}
\affiliation{School of Science, Inner Mongolia University of Science \& Technology, Baotou, Inner Mongolia Autonomous Region 014010, China}
\affiliation{School of Mathematics and Physics, Jiangsu University of Technology, Changzhou, Jiangsu 213001, China}

\date{\today}

\begin{abstract}
In high energy heavy-ion collisions, the final anisotropic flow coefficients and their corresponding event-plane correlations are considered as the medium evolutional response to the initial geometrical eccentricities and their corresponding participant-plane correlations. We formulate a systematic theoretical analysis to study the hydrodynamical responses concerning higher order effects in Pb+Pb collisions at $\sqrt{s_\text{NN}}=2.76$\,TeV by using Monte Carlo Glauber (MC-Glauber) model. To further understand the transformations of the initial participant-plane correlation, we construct a new set of events which randomize the directions of the initial participant planes of the original events. Our results indicate that the final strong event-plane correlations are mainly transformed from the large initial eccentricities, rather than the strong participant-plane correlations. However, the large flow coefficients and the discrepancies between the flow coefficients calculated by the single-shot and event-by-event simulations in peripheral collisions are relevant to those strong initial participant-plane correlations.
\end{abstract}

\maketitle

\section{Introduction}

It is widely accepted that the extreme hot and dense matter created in the heavy ion collisions
at the Relativistic Heavy Ion Collider\,(RHIC) and the Large Hadron Collider\,(LHC) show event-by-event fluctuations.
Due to the fluctuating positions of the nucleons inside the colliding nuclei and
the quantum fluctuations of the quark and gluon fields inside those nucleons,
the initial quantities of each event show fluctuations in size, shape and magnitude.
With the following dynamical evolutions, the initial geometrical fluctuations gradually transform into the anisotropies in the final momentum distribution of the emitted particles.
To understand these transformations, four quantities are defined in the transverse plane: the initial eccentricity coefficients $\varepsilon_n$ and their corresponding participant-plane azimuthal angles $\Phi_n$,
the final particle momentum flow coefficients $v_n$ and their corresponding event-plane flow angles $\Psi_n$\,\cite{PhysRevD.46.229,ZPhysC.70.665,RelativisticHeavyIonPhysics}.
In this definition, $\{v_n,\Psi_n\}$ are considered as the medium responses of those $\{\varepsilon_n,\Phi_n\}$, {\it i.e.}, the initial $\{\varepsilon_n,\Phi_n\}$ are encoded in the final $\{v_n,\Psi_n\}$ during the evolution process\,\cite{AHEP.2017.1485353}.
Therefore, studying the transformation from $\{v_n,\Psi_n\}$ to $\{\varepsilon_n,\Phi_n\}$ can help us understand the initial geometry structure of the system in heavy-ion collisions.

In the past few years, many efforts have been devoted to explore the transformation
from the initial $\{\varepsilon_n,\Phi_n\}$ to the final $\{v_n,\Psi_n\}$.
In Ref.\,\cite{AnnuRevNuclPartSci.57.205,PhysRevC.82.064903,PhysRevC.85.024908,NuclPhysA.931.959,JPhysG.41.063102,PhysLettB.744.82,PhysRevC.93.014909,AHEP.2018.8347408,PhysRevC.98.044908},
the authors study the transformation from the initial state eccentricities to the final state flows.
While in Ref.\,\cite{PhysLettB.717.261,EurPhysJC.73.2510,NuclPhysA.904.365c},
the authors study the transformation from the initial state participant-plane correlations to the final-state event-plane correlations.
Recently, people realize that study four quantities together, {\it i.e.}, initial state eccentricities, participant-plane correlations, final state flows and participant-plane correlations, will provide new information for further understanding the transformation effect of the system evolution process\,\cite{JPhysG.41.124003,PhysRevC.94.014906,AHEP.2018.6914627,ChinPhysC.42.042001}.
In addition, flow harmonic coefficients up to sixth-order with their associated event-plane correlations
for Pb+Pb collisions at $\sqrt{s_\text{NN}}=2.76$ TeV have been published and discussed in LHC experiment\,
\cite{PhysRevLett.107.032301,PhysLettB.708.249,EurPhysJC.72.2012,NuclPhysA.904.511c,NuclPhysA.910.276,IntJModPhysA.29.1430044,PhysRevC.92.034903,PhysLettB.773.68,JHEP.07.103}.
Therefore, a complete and systematic theoretical analysis concerning the high-order harmonic coefficients with their plane correlations is needed.

Theoretically, hydrodynamics is an efficient tool in describing the evolution of the system
from quark-gluon plasma (QGP) phase to hadron gas (HG) phase\,\cite{IntJModPhysE.24.1530010,RepProgPhys.81.046001}.
In early experiments, due to the limited particle emissions in each event,
data of millions of events are needed to obtain the meaningful statistical results.
Correspondingly, a meaningful theoretical analysis involves event-by-event hydrodynamical calculations,
{\it i.e.}, the initial energy density profile for each simulated collision are evolved individually, which is extremely resource demanding and time consuming\,\cite{NuclPhysA.982.927}.
Besides, the final results are calculated from the particles of all the simulated events.
In an early analysis, due to the limited data processing capability of the computer hardware, single-shot simulation was widely used to simplify the problem.
In this way, one can average over the multiple fluctuating profiles to obtain a single relatively smooth initial profile, and then feed it into the hydrodynamics\,\cite{ComputPhysCommun.199.61}.
Although the latter method ignores the detailed individual fluctuations inside the initial distributions of the thermal quantities, one can see that it captures some characteristics of the created hot matter to some certain extent\,\cite{JPhysG.42.045105}.
For example, at small transverse momentum region (0-1.5\,GeV/$c$) of particles, the final state spectrum and the elliptic flow of the single-shot hydrodynamical evolutions are consistent with the ones obtained by  event-by-event hydrodynamical evolutions. However, for large transverse momentum region ($\geq$1.5\,GeV$/c$) of particles, the results of the spectrum and the elliptic flow of single-shot evolution are higher than the ones obtained by event-by-event hydrodynamical evolutions.
Therefore, applying both two kinds of simulations in the study of the transformation from $\{\varepsilon_n,\Phi_n\}$ to $\{v_n,\Psi_n\}$
are valueable for us to understand the hydrodynamical transformations.

The remaining of this paper is arranged as follows.
In section II, we review the model and the main methodology.
In section III, we show our results and make some discussions.
We start from studying the equivalence of the single-shot and event-by-event simulations through analysing initial eccentricities $\varepsilon_2$ to $\varepsilon_6$ and their corresponding final anisotropic flow coefficients $v_2$ to $v_6$ of the most central, the middle and the peripheral centralities with different Gaussian width $\sigma$ in MC-Glauber model at Pb+Pb $\sqrt{s_\text{NN}}=2.76$ TeV collisions.
Then, we show the event-by-event hydrodynamical transformation of the two-plane correlations up to the ninth-order from the initial state to the final state.
We find that, due to the strong initial correlations and the nonlinear hydrodynamical evolution,
the equivalence of the results from two simulation methods breaks down in the peripheral collisions.
At the same time, the results of the plane correlations imply dynamical rotations of the plane during the evolution.
To further understand the hydrodynamical response of the plane correlations,
We construct a new set of events on the basis of the original events for 70-80\% centrality interval with $\sigma=0.3$ fm.
Compared with the results of the original event set, a detailed nonlinear hydrodynamical effect, which concerns the cross-effects including the initial eccentricities, their corresponding planes, the final flows and their corresponding planes, is further discussed.
In Section IV, we give summary and conclusions.

\section{Model and Methodology}

In this work,
we extract the MC-Glauber part (initial condition) and VISHNEW part (hydrodynamical evolution)
from iEBE-VISHNU code package which developed by Chun Shen {\textit{et al.}}\,\cite{ComputPhysCommun.199.61},
and combine with our Monte Carlo final particle collector code to
calculate pion transverse spectrum and anisotropic flow of Pb+Pb $\sqrt{s_\text{NN}}=2.76$ TeV collisions at LHC\,\cite{AHEP.2015.846154,NuclSciTech.27.147,ChinPhysC.41.084102}.
The model parameters are taken according as Ref.\,\cite{PhysLettB.707.151} reported.

We generate fluctuating initial conditions
for Pb+Pb collisions at $\sqrt{s_\text{NN}}=2.76$ TeV using MC-Glauber model.
The total entropy density produced in the transverse plane after thermalization
is taken to be a mixture of the wounded nucleon (WN) and binary collision (BC) density profiles,
\begin{equation}
s(\vec{r}_\perp)=\mathcal{K}_s\left[{\frac{1-\alpha}{2}\,\mathrm{WN}(\vec{r}_\perp)+\alpha\,\mathrm{BC}(\vec{r}_\perp)}\right]
\end{equation}
where
\begin{equation}
\begin{aligned}
&\mathrm{BC}(\vec{r}_\perp)=\sum_{(i,j)\in \mathrm{pairs}}{\frac{1}{2\pi\sigma^2}e^{-\frac{|\vec{r}_\perp-\vec{R}_{ij,\perp}|^2}{2\sigma^2}}},\\
&\mathrm{WN}(\vec{r}_\perp)=\sum_{i\in \mathrm{wounded}}{\frac{1}{2\pi\sigma^2}e^{-\frac{|\vec{r}_\perp-\vec{r}_{i,\perp}|^2}{2\sigma^2}}}.
\end{aligned}
\label{BC-and-WN}
\end{equation}
$\vec{r}_{\perp}$ is the transverse spatial coordinates,
$\vec{R}_{ij,\perp}$ and $\vec{r}_{i,\perp}$ are the collision spot of the nucleon pair and the coordinate of the wounded nucleon respectively.
$\sigma$ is the Gaussian width, which is taken to be the same for both WN and BC parts.
$\alpha$ is the binary mixing parameter, which is tuned to reproduce its observed dependence on collision centrality,
and $\mathcal{K}_s$ is the overall normalization factor, which is tuned to reproduce the measured final charged multiplicity in the most central collisions.

With the help of equation of state\,(EOS), the entropy density profile $s(\vec{r}_\perp)$ can be converted to the energy density form $e(\vec{r}_\perp)$.
In this way, the initial transverse geometry eccentricity can be defined as\,\cite{PhysRevLett.98.242302}
\begin{equation}
\begin{aligned}
&\varepsilon_1 e^{i\Phi_1}=-\frac{\int dxdyr^3e^{i\phi} e(x,y)}{\int dxdyr^3 e(x,y)},\\
&\varepsilon_n e^{in\Phi_n}=-\frac{\int dxdyr^ne^{in\phi} e(x,y)}{\int dxdyr^n e(x,y)},~~(n>1)
\end{aligned}
\label{Eq-ecc}
\end{equation}
where $\varepsilon_n$ is the $n$-th order harmonic eccentricity coefficient,
and $\Phi_n$ is the angle of the corresponding eccentricity plane,
which points to the minor axis of the plane and is also known as participant-plane.
One need to notice that the coordinate system of the energy density profile $e(x,y)$ here
is shifted to the center of mass of the participating nucleons.

The following evolution process is based on a (2 + 1) dimensional viscous hydrodynamic model
with longitudinal boost-invariance, describing numerically the transverse evolution
of the created matter in high energy heavy-ion collision near midrapidity.
The initial profile is hydrodynamically evolved from the proper time 0.6 fm/$c$ with EOS s95p-PCE\,\cite{PhysRevC.82.054904}
which combines numerical lattice QCD results at high temperatures
with hadron resonance gas at low temperatures\,\cite{NuclPhysA.837.26}
and implements chemical freeze-out at the temperature 165 MeV.
Thus, we only concentrate on the pions of the final state hadrons in this work.
We convert hydrodynamic outputs into Bose-Einstein distributions for pions
along an isothermal decoupling surface with temperature 120 MeV using the viscous corrected Cooper-Frye description\,\cite{PhysRevD.10.186}.
The shear viscous coefficient $\eta/s$ is taken as 0.08 for both QGP and HG phase, and the bulk viscosity is not considered for simplicity.
For simplicity, the hadronic rescattering in the dilute hadron gas and the resonance decays are not considered, since that they should not have much influence on
the results for the charged particle flow coefficients and their event-plane correlations\,\cite{PhysRevC.82.064903,PhysLettB.717.261,PhysRevC.81.044906}.

The final transverse momentum anisotropic flow of the particles are defined by
\begin{equation}
\begin{aligned}
&v_n e^{in\Psi_n}=\frac{\int d\eta \int p_Tdp_Td\phi_p e^{in\phi_p}\frac{dN}{p_Tdp_Td\phi_p d\eta}}
                       {\int d\eta \int p_Tdp_Td\phi_p \frac{dN}{p_Tdp_Td\phi_pd\eta}},
\end{aligned}
\label{Eq-flow}
\end{equation}
where $v_n$ is the $n$-th order harmonic flow coefficient,
and $\Psi_n$ corresponds to the same order flow plane which is also known as the $n$-th order event-plane.
For the calculation of $v_1$, the authors of Ref.\,\cite{PhysRevC.85.024908,PhysRevLett.106.102301} argued that
an average with equal weighting such as in Eq.\,(\ref{Eq-flow}) is not appropriate,
since that $v_1$ changes sign as a function of transverse momentum.
Therefore, we do not take $v_1$ and $\Phi_1$ into consideration here.
In addition, we only calculate final state flow coefficients and participant-plane correlations for pions, as pions are the most abundant particles at the final state.
We take the same pseudorapidity range $0.5<|\eta|<2.5$ and $p_T > 0.5$ GeV,
as those used in the ATLAS experimental analysis\,\cite{NuclPhysA.910.276}.

In order to understand the evolutional transformation from $\varepsilon_n$ to $v_n$,
we compare two kinds of simulations, {\it i.e.}, single-shot and event-by-event simulations.
To accurately characterize the eccentricity for a certain order in single-shot simulation,
each initial entropy density distribution need to be rotated according to the participant-plane of concerned order,
{\it i.e.}, we need to rotate the initial conditions to align their orientations $\Phi_n$ before averaging.
In this way, the resulting averaged profiles are different for different $n$.
For example, aligning $\Phi_2$ for all events will result in an ellipse-shaped averaged profile, while aligning
$\Phi_3$ for all events will result in a triangular-shaped averaged profile.
Therefore, each order average profile only highlights the eccentricity information of this certain order and only responsible for the estimation of the flow of this specific order.

\section{Results and Discussions}

\begin{figure*}[htb]
\begin{center}
  \includegraphics[scale=0.5]{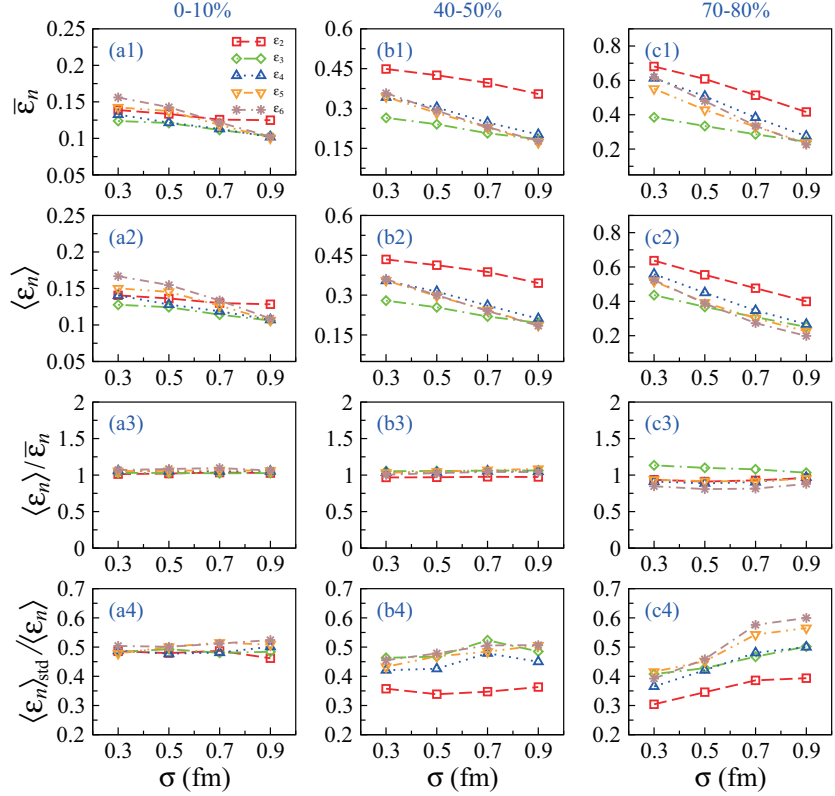}
  \caption{(Color online)
  Top row: the eccentricity calculated by each order rotate average single-shot initial conditions versus $\sigma$.
  Second row: the assemble average of the event-by-event eccentricities versus $\sigma$.
  Third row: the ratio of the assemble average of the event-by-event eccentricity to the rotate average single-shot eccentricity.
  Bottom row: the coefficient of variance $\langle\varepsilon_n\rangle_{\rm std}/\langle\varepsilon_n\rangle$
  calculated by event-by-event eccentricities versus $\sigma$.}
  \label{ebe-ecc}
\end{center}
\end{figure*}

In Fig.\,\ref{ebe-ecc}, we compare the eccentricity of rotate-average single-shot $\bar{\varepsilon}_n$
and event-by-event $\langle \varepsilon_n\rangle$ versus $\sigma$
for 1200 events of $\sqrt{s_\text{NN}}=2.76$ TeV Pb+Pb collisions in 0-10\%, 40-50\% and 70-80\% centrality intervals, respectively.
For each event, we sample $10^8$ pions from its hydrodynamical freeze-out hypersurface.
Here, $\langle\cdots\rangle$ represents a final state ensemble average for all concerned events.
The harmonic order is taken from the second up to the sixth.
The connected lines are used to easily distinguish data from different orders.
Both $\bar{\varepsilon}_n$ and $\langle \varepsilon_n\rangle$ show similar trends with $\sigma$ for three centrality intervals.
They increase with the decreasing collision centralities.
Due to the highly fluctuating positions of the nucleons inside the colliding nuclei,
the odd order participant nucleon eccentricities are obviously non-zero in both single-shot and event-by-event calculations,
even for the very central collisions.
We see $\bar{\varepsilon}_n\cong\langle\varepsilon_n\rangle$ for all the cases as shown in the third row of Fig.\,\ref{ebe-ecc}.
The average similarities $\sum_{i}|\bar{\varepsilon}_{ni}-\langle\varepsilon_{ni}\rangle|/4\langle\varepsilon_{ni}\rangle$ for three centralities are 4.0\%, 3.9\%, 7.4\%, respectively, where the subscript $i$ represents the cases of $\sigma=0.3$, 0.5, 0.7 and 0.9.
It indicates that the rotate-average operation for single-shot calculation could carry the eccentricity information of its corresponding order,
although it smooths out the fluctuating structure inside each single event as well as weakens other order information.
Therefore, one can quantitatively understand the initial geometry characteristics of the collision source through a series of rotate-average single-shot initial profiles.
We also calculate the coefficient of variance $\langle\varepsilon_n\rangle_{\text{std}}/\langle\varepsilon_n\rangle$
to characterize the concentration degree of $\varepsilon_n$ of these cases, see the bottom row.
The subscript ``$\text{std}$'' represents the standard deviation.
It shows that the initial eccentricities for peripheral collisions are more concentrate than the central collision ones when $\sigma<0.7$ fm, because of the strong geometry limitation of the initial participant zone for the peripheral collisions.
This concentration gradually breaks for large $\sigma$ case in peripheral collisions, especially for higher order eccentricities.
For these cases, large $\sigma$ may cause the packet of the fluctuating distributions overlap each other which modifies the fluctuations of the initial profiles.

\begin{figure*}[htb]
\begin{center}
  \includegraphics[scale=0.5]{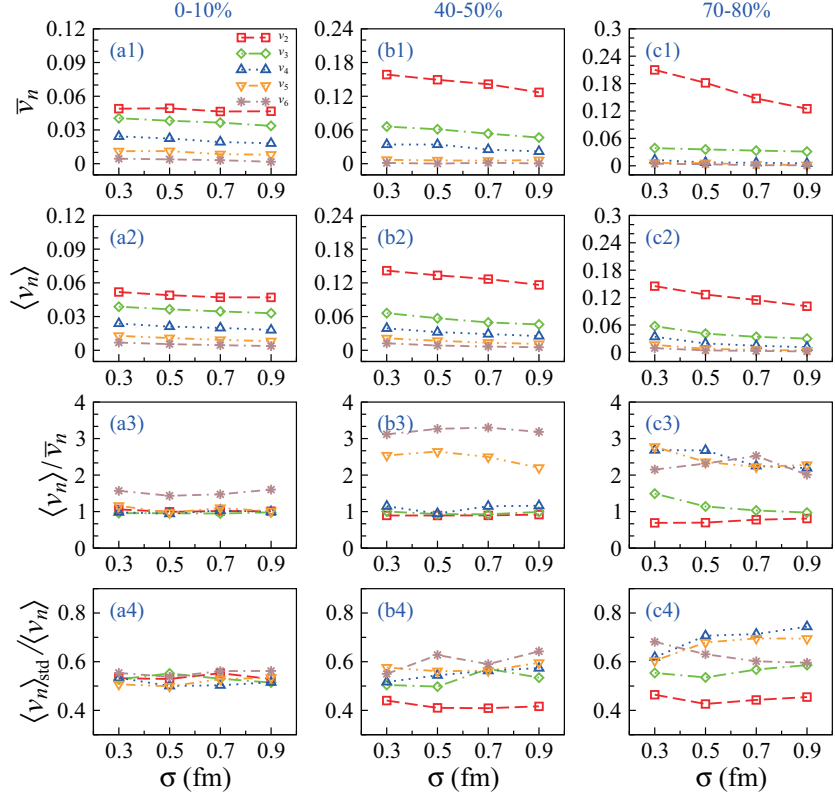}
  \caption{(Color online) Same to Fig.\,\ref{ebe-ecc}, but for the final state flow results.}
  \label{ebe-flow}
\end{center}
\end{figure*}

The corresponding final flow results are shown in Fig.\,\ref{ebe-flow}.
Unlike the initial eccentricities, $\bar{v}_n\cong\langle v_n\rangle$ are not shown for all the considered cases.
It is only valid for the most central collisions and the large $\sigma$ cases in peripheral collisions,
{\it i.e.}, the cases with small initial eccentricities,
which indicates that one can equivalently use simple single-shot hydrodynamical analysis to substitute the tedious event-by-event one for these cases.
This equivalence gradually breaks down with the decreasing collision centrality, and the average similarities $\sum_{i}|\bar{v}_{ni}-\langle v_{ni}\rangle|/4\langle v_{ni}\rangle$ for three centralities are 12.1\%, 31.2\%, 37.1\%, respectively, where the subscript $i$ represents the cases of $\sigma=0.3$, 0.5, 0.7 and 0.9.
In small $\sigma$ peripheral collisions, $\bar{v}_2$ is obviously higher than $\langle v_2\rangle$,
meanwhile other order $\bar{v}_n$ and $\langle v_n \rangle$ also show discrepancies to different extents.
Considering that a certain order single-shot initial profile can only capture the geometrical eccentricity of its own order,
we believe that the differences between $\bar{v}_n$ and $\langle v_n\rangle$ are caused by a combination effect from three aspects.
One is the relatively large initial $\varepsilon_n$ in some events ({\it e.g.} most of the events with small $\sigma$ for peripheral collisions) which will affect the results of final flow coefficients significantly.
Another one is the absence of the information in single-shot simulations which is related to the fluctuations from other orders at the initial states.
The third one is the correlation effect produced by the interactions between different order participant planes during the evolution process, which indicates a nonlinear transformation from $\varepsilon_n$ to $v_n$.
This point can also be inferred from the bottom three panels of Fig.\,\ref{ebe-ecc} and Fig.\,\ref{ebe-flow}.
We see that the variation of concentration degree of $v_n$ and $\varepsilon_n$ with $\sigma$ show obvious difference, and the values $v_n$ are wholly larger than the values of $\varepsilon_n$.
It means that the concentration degree of final flow coefficients for the peripheral collisions are more discrete than the central collisions, except for $v_2$ which is strictly constrained by the collision geometry at the initial state.

\begin{figure}
\begin{center}
  \includegraphics[scale=0.45]{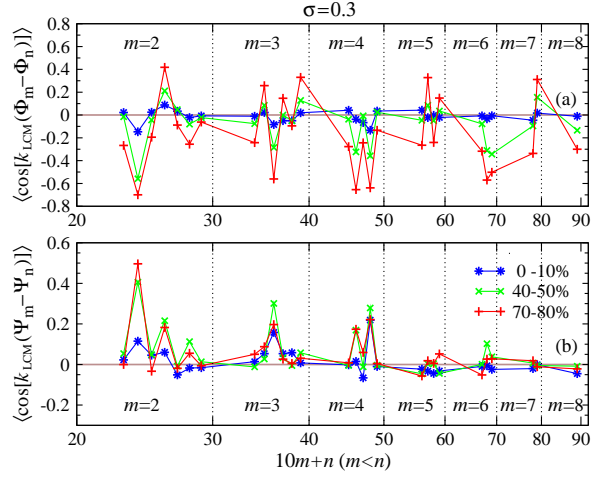}
  \caption{(Color online) Top panel: Two-plane correlations $\langle\cos[k_{\text{LCM}}(\Phi_m-\Phi_n)]\rangle$ between pairs of participant-plane angles $\Phi_n$ and $\Phi_m$. Bottom panel: for the corresponding correlations between final state event-plane angles $\Psi_n$ and $\Psi_m$.}
  \label{corrpsi-corrphi}
\end{center}
\end{figure}

In Fig.\,\ref{ebe-ecc} and Fig.\,\ref{ebe-flow}, we only focus on the coefficients of the initial geometrical eccentricity and the final momentum flow.
However, a complete understanding of the entire fluctuations should include their corresponding angle information.
The authors of Ref.\,\cite{PhysRevC.83.044908,EurPhysJC.73.2558} also pointed out that correlations between the event angles of different orders
(hereafter referred to as ``final event-plane correlation'' or ``final $\Psi_m$-$\Psi_n$ correlation'')
can yield valuable additional insights into the initial conditions.
Therefore, we further investigate the angle correlations between different order planes
of the eccentricity participant-plane
(hereafter referred to as ``initial participant-plane correlation''
or ``initial $\Phi_m$-$\Phi_n$ correlation'') and the flow event-plane.
Considering the limitations of single-shot calculation, {\it i.e.}, each rotate-averaged initial distribution only highlights a certain order eccentricity but ignores the informations from other orders, the followings calculations are only performed for event-by-event hydrodynamical analysis.
The initial participant-plane correlation $\langle\cos[k_{\text{LCM}}(\Phi_m-\Phi_n)]\rangle$
and the final event-plane correlation $\langle\cos[k_{\text{LCM}}(\Psi_m-\Psi_n)]\rangle$
for different $m$ and $n$ are plotted in Fig.\,\ref{corrpsi-corrphi}, where $k_{\text{LCM}}$ is the least common multiple (LCM) of $n$ and $m$.
In the rest of this paper, we take the case of $\sigma=0.3$ fm to discuss, because it gives relatively large values for most of the concerned $\varepsilon_n$ and $v_n$ in Fig.\,\ref{ebe-ecc} and Fig.\,\ref{ebe-flow}.
The orders are taken up to ninth.
The numerical errors, which may caused by varies of the fluctuating profile, are checked by jackknife resampling approach.
To make the data clear to read, we take the horizontal axis to be logarithmic and plot connected solid lines between two nearest sperate symbols.
In Fig.\,\ref{corrpsi-corrphi}\,(a), some $\langle\cos[k_{\text{LCM}}(\Phi_m-\Phi_n)]\rangle$ show positive values but others show negative values.
When evolving to the final state, almost all $\langle\cos[k_{\text{LCM}}(\Psi_m-\Psi_n)]\rangle$ become positive.
The result implies the existence of the dynamical rotations of all the event-planes during the evolutions, even for 0-10\% collisions. These are also shown in Ref.\,\cite{PhysLettB.717.261}.
It once more reflects the nonlinear characteristic of the hydrodynamics.
Besides, with the decreases of the collision centrality, {\it i.e.}, from central to peripheral collisions, the interactions between different order participant planes become more stronger,
see that $\langle\cos[k_{\text{LCM}}(\Psi_m-\Psi_n)]\rangle$ at 70-80\% collisions show more deviation from the horizontal zero base line.
Meanwhile, several correlations exhibit dramatically different centrality dependencies
for the initial participant-plane and the final event-plane angles, see the correlations between second and sixth order, second and eighth order, third and sixth order, and fourth and eighth order, {\it etc.}, which also implies in Ref.\,\cite{PhysLettB.717.261}.
Moreover, some $\Phi_m$-$\Phi_n$ correlations, where $n$ is an integer multiple of $m$ ({\it e.g.}, $\Phi_2$-$\Phi_4$, $\Phi_2$-$\Phi_6$, $\Phi_2$-$\Phi_8$, $\Phi_3$-$\Phi_6$, $\Phi_4-\Phi_8$ correlations),
or $n$ and $m$ have common divisor ({\it e.g.}, $\Phi_4$-$\Phi_6$ correlation), exhibit more stronger than the other order correlations.
And these correlations can all survive until the final state, even for the most central collisions.
Considering above relations between $m$ and $n$, we believe these strong initial state and final state correlations are due to the strong geometrical symmetry limitations between these certain orders.

We now give a short summary here.
In small $\sigma$ peripheral collisions, Fig.\,\ref{ebe-ecc} shows large eccentricities in the participant-plane of the initial state due to the fluctuating effects, and Fig.\,\ref{corrpsi-corrphi}\,(a) exhibits strong angle correlations between two participant-planes of different orders.
Then, the final state results in Fig.\,\ref{ebe-flow} and Fig.\,\ref{corrpsi-corrphi} together reveal the interactions between different order planes yield dynamical rotation of each order plane during the evolution process,
and we interpret this as the nonlinear behaviour of the hydrodynamics.
In calculations, hydrodynamics is just treated as a mathematical tool, the evolution equations are the same for all the collision cases.
Therefore, it is not easy to distinguish which factor plays the most important role on the final measurable strong event-plane correlations for peripheral collisions:
the large initial eccentricities, the strong initial participant-plane correlations or the nonlinear hydrodynamical effects during the evolution.
We now only focus on the hydrodynamical transformations of the initial participant-plane correlations.
To further understand it, inspired by the method proposed in Ref.\,\cite{JPhysG.41.015103}, we create a new set of events on the basis of the previous original events,
{\it i.e.}, we keep the eccentricities for different orders but assign their participant-planes to a random rotation.
The main ideas are explained as follows.

In the system of polar coordinates, $e(x,y)$ can be decomposed in terms of the azimuthal angle $\phi=\arctan(y/x)$ by Fourier series,
\begin{equation}
e(r,\phi)=\,e_0(r)+2\sum_{n=1}^{\infty}\left[{e_n^c(r)\cos(n\phi)+e_n^s(r)\sin(n\phi)}\right].
\end{equation}
Inserting it into Eq.(\ref{Eq-ecc}), $\varepsilon_n$ can be expressed as
\begin{equation}
\begin{aligned}
&\varepsilon_1 e^{i\Phi_1}
=-\frac{\int dr r^4e_1^c(r)}{\int dr r^4 e_0(r)}-i\frac{\int dr r^4e_1^s(r)}{\int dr r^4 e_0(r)},\\
&\varepsilon_n e^{in\Phi_n}
=-\frac{\int dr r^{n+1}e_n^c(r)}{\int dr r^{n+1} e_0(r)}-i\frac{\int dr r^{n+1}e_n^s(r)}{\int dr r^{n+1} e_0(r)},\,(n>1).
\end{aligned}
\end{equation}
Here, each pair of the coefficients $e_n^c(r)$ and $e_n^s(r)$ only contribute to the $\varepsilon_n$ of the same order $n$.
Based on this idea, one can hold $e_n^{c,s}(r)$ and then rotate each $n\phi$ to another angle $n(\phi-\Theta_n')$,
therefore arrive at a new $e'(r,\phi)$,
\begin{equation}
\begin{aligned}
e'(r,\phi)
=&\,e_0(r)+2\sum_{n=1}^{\infty}\left\{e_n^c(r)\cos[n(\phi\,-\,\Theta_n')]\right.\\
                              &\left.+\,e_n^s(r)\sin[n(\phi\,-\,\Theta_n')]\right\}.
\end{aligned}
\end{equation}
Through a short derivation, $\{\varepsilon_n',\,\Phi_n'\}$ for $e'(r,\phi)$ can be easily expressed as
\begin{equation}
\begin{aligned}
&\varepsilon_n'=\varepsilon_n,\\
&\Phi_n'=\Phi_n+\Theta_n',
\end{aligned}
\end{equation}
which means this new initial profile carries the same eccentricities but different participant-planes for each order.

\begin{figure*}[htb]
\begin{center}
  \includegraphics[width=0.8\textwidth]{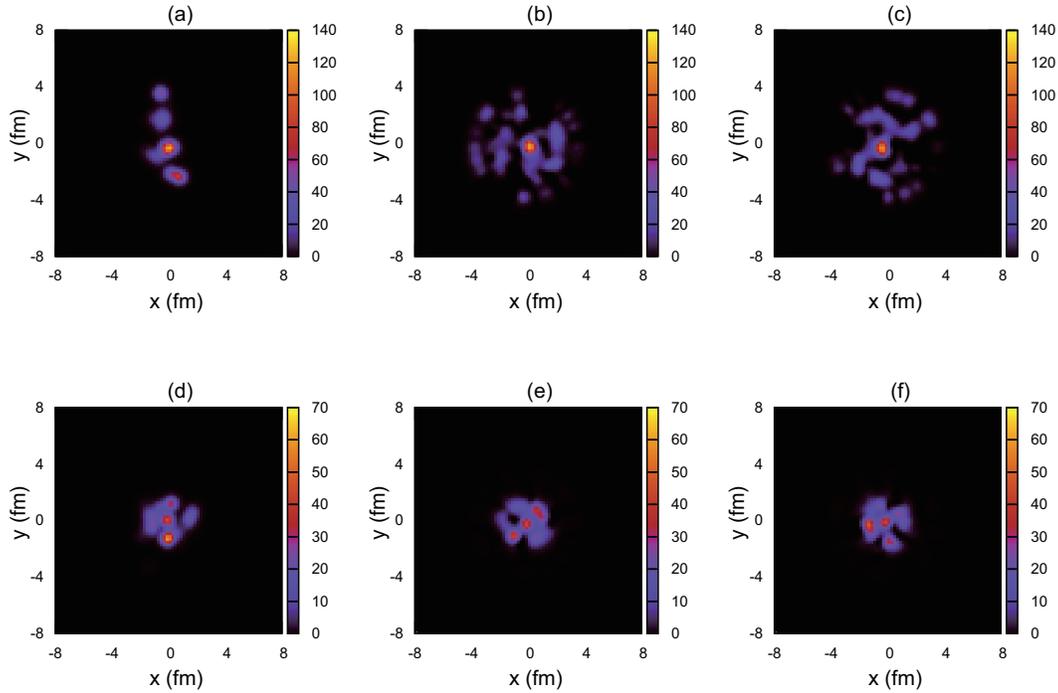}
  \caption{(Color online) A comparison of energy density distribution in transverse plane of the randomly plane event and the original event. (a) and (d) are two original events. (b) and (c) are two randomly plane events on the basis of event (a). (e) and (f) are two ones on the basis of event (d).}
  \label{event-rotate}
\end{center}
\end{figure*}

Based on the theory above, we create a new set of events through random sampling the first fifty orders of $\Theta_n'$ (hereafter are labelled as $\Theta_n'\,(n\leq50)$)
on the basis of the original event sets which we have used in 70-80\% with $\sigma=0.3$ fm simulations.
To make this clear, we show a comparison of 2D energy density distribution in transverse plane of the randomly plane event and the original event in Fig.\,\ref{event-rotate}.
Fig.\,\ref{event-rotate}\,(b) and (c) are two randomized $\Theta_n'\,(n\leq50)$ events on the basis of the original event shown in Fig.\,\ref{event-rotate}\,(a).
Fig.\,\ref{event-rotate}\,(e) and (f) are on the basis of the original event shown in Fig.\,\ref{event-rotate}\,(d).
We see that although the event (a), (b), (c) have the same eccentricities, they look totally different, and so as the event (d), (e), (f).
However, no matter how much the planes be rotated, the matter in the center of the plane will always stay at their original position.
Also, the range of the distributions for these randomly plane events can not exceed the maximum radial positions which are decided by their corresponding original events.
\begin{figure}
\begin{center}
  \includegraphics[scale=0.45]{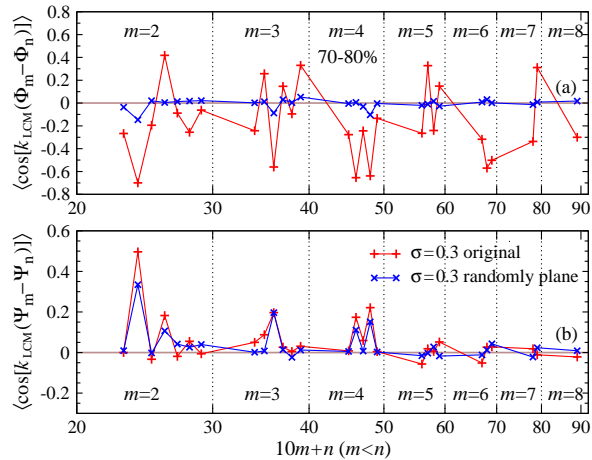}
  \caption{(Color online) Similar to Fig.\,\ref{corrpsi-corrphi}, but for comparing the results calculated by the randomize $\Psi_n'\,(n\leq50)$ events with their corresponding ones from the original events.}
  \label{corrpsi-corrphi-rotation}
\end{center}
\end{figure}

Now, we compare their initial $\Phi_m$-$\Phi_n$ and final $\Psi_m$-$\Psi_n$ correlations for each corresponding $m$ and $n$ orders in 70-80\% centrality with $\sigma=0.3$ fm collision.
The results are shown in Fig.\,\ref{corrpsi-corrphi-rotation}.
Here, the data for the original events are exactly the one plotted in Fig.\,\ref{corrpsi-corrphi},
and the one for the randomized $\Theta_n'\,(n\leq50)$ events are labelled as ``randomly plane''.
After randomly rotating the participant-plane,
one can see that strong initial participant plane correlations are almost eliminited,
except for some pairs of different order numbers in which one is a multiple of the other, {\it i.e.}, $\Phi_2$-$\Phi_4$, $\Phi_3$-$\Phi_6$ and $\Phi_4$-$\Phi_8$ correlations.
The small non-zero values of these correlations indicate that these participant planes are still weakly correlated to each other.
We believe these correlations can be barely further weaken because of an unavoidable combination effect which comes from the frequency doubling problem in the Fourier expansion
and the numerical discrete problem during the transformation between Cartesian and polar coordinate systems.
Meanwhile, compared with the strong correlations of the original events, these weak values are small enough, so that they have little effect on the conclusions we are about to reach.
From the final state results in Fig.\,\ref{corrpsi-corrphi-rotation}\,(b), two sets of data show similar trends.
This means that, no matter how weak the initial state participant-plane correlations are, the final state event-plane correlations are still strong
for the peripheral collisions which carry large initial eccentricities.
As shown in Fig.\,\ref{ebe-ecc} and Fig.\,\ref{corrpsi-corrphi}, the final state event-plane correlations are weak for the central collisions which carry small eccentricities and weak participant-plane correlations.
Therefore, we conclude that the strong final event-plane correlations
mainly come from the hydrodynamical transformation of the large eccentricities at the initial state,
rather than their strong initial state correlations, although in fact that the large eccentricities are definitely accompanied by strong initial state participant-plane correlations in a reasonable simulated event in high energy heavy-ion collisions.
Therefore, one can not qualitatively get the corresponding initial participant-plane correlations just from the final event-plane correlation data.

So far, we have separately studied the hydrodynamical transformation from $\varepsilon_n$ to $v_n$,
as well as the one from $\Phi_m$-$\Phi_n$ to $\Psi_m$-$\Psi_n$ correlations.
In fact, for any single fluctuating event, its eccentricities and the corresponding participant planes are intricately entangled.
This entanglement is mainly from three aspects: the entanglement between the eccentricity and the eccentricity plane from the same order ({\it i.e.}, the correlation between $v_2$ and $\Phi_2$),
the entanglement between the two or more harmonic coefficients from the different orders especially for the numbers of the orders having multiple relations ({\it i.e.}, the correlation between $v_2$ and $v_4$),
and the entanglement between the eccentricity from one order and the harmonic planes from other orders ({\it i.e.}, the correlation between $v_2$ and $\Phi_3$).
Our discussions above concern the harmonic order up to the ninth.
Besides, to cancel out the initial state two-plane correlations between the lower and higher orders,
the randomization of $\Theta_n'$ for each initial profile is operated even to the fiftieth order.
For further understanding this entanglement problem, our next discussions only focus on two cases:
correlations between the second and third orders, and correlations between the second and fourth orders in 70-80\% collision.
The data of 70-80\% collision in Fig.\,\ref{corrpsi-corrphi} show that
both the initial correlations of $\Phi_2$-$\Phi_3$ and $\Phi_2$-$\Phi_4$ are negative,
but the final correlation of $\Psi_2$-$\Psi_3$ approaches to zero and the correlation of $\Psi_2$-$\Psi_4$ becomes a ``large'' positive number.
The explanations for correlations between the second and third orders, and correlations between the second and fourth orders are different.
The former one can be attributed by the inside fluctuations of the matter,
but the latter one is due to the geometrical multiple frequency effect between two order numbers.
Therefore, we take this two typical correlations for the following discussions.
\begin{figure*}[htb]
\begin{center}
  \includegraphics[scale=0.53]{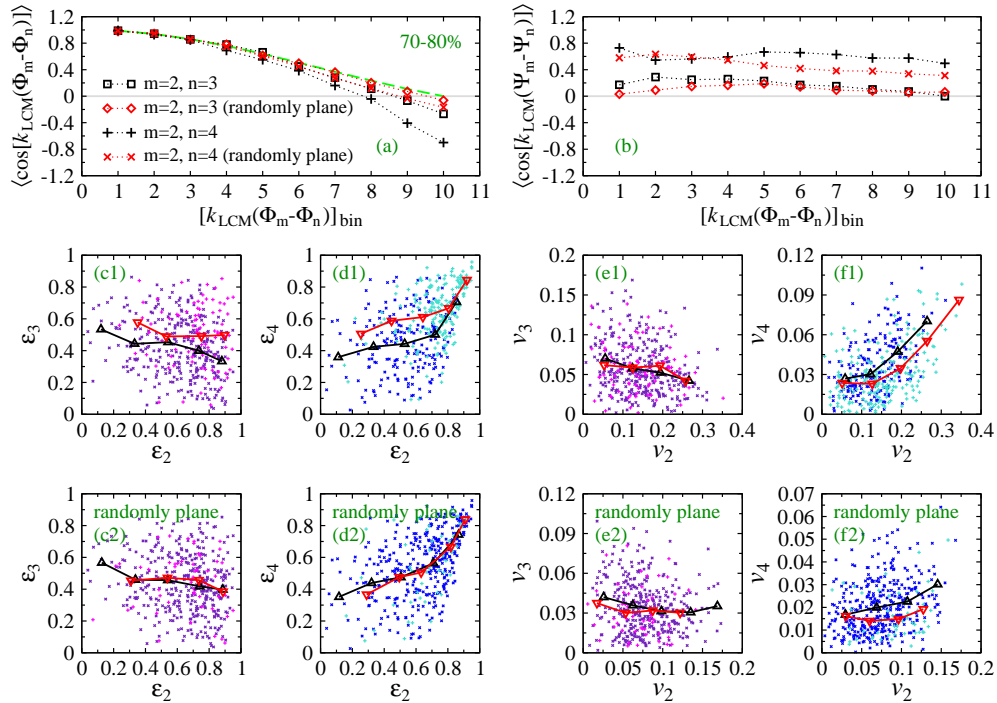}
  \caption{(Color online)
  Top row: two-plane correlations $\langle\cos[k_{\text{LCM}}(\Phi_m-\Phi_n)]\rangle$ and $\langle\cos[k_{\text{LCM}}(\Psi_m-\Psi_n)]\rangle$
  versus $[k_{\text{LCM}}(\Phi_m-\Phi_n)]_{\text{bin}}$ respectively, see text for details.
  The green dashed line represent the cases have no correlations between $\Phi_m$ and $\Phi_n$.
  Middle row: the scatters of $\varepsilon_2$-$\varepsilon_3$, $\varepsilon_2$-$\varepsilon_4$, $v_2$-$v_3$ and $v_2$-$v_4$
  for the original events. The light color (magenta and cyan) represents the events which belong to the 10th bin, and the dark color (dark purple and dark blue) represents the events which belong to the other bins.
  Bottom row: similar to the middle row, but for $\Theta_n'\,(n\leq50)$ randomized events.}
\label{sort1}
\end{center}
\end{figure*}

We proceed a further hydrodynamical transformation study from $\Phi_m$-$\Phi_n$
to $\Psi_m$-$\Psi_n$ correlation for $(m,\,n)$ equals (2,\,3) and (2,\,4).
The results are shown in the top row of Fig.\,\ref{sort1}.
The initial $\Phi_2$-$\Phi_3$ and $\Phi_2$-$\Phi_4$ correlations are calculated from two sets of initial conditions,
the original event set and the $\Theta_n'\,(n\leq50)$ randomized event set.
We separate the range of $[k_{\text{LCM}}(\Phi_m-\Phi_n)]$, which is from 0 to $\pi$, into 10 bins,
and mark the bin numbers in the horizonal axis sequently.
The vertical coordinate of each data point is the average of $\cos[k_{\text{LCM}}(\Phi_m-\Phi_n)]$ [Fig.\,\ref{sort1}-(a)]
and $\cos[k_{\text{LCM}}(\Psi_m-\Psi_n)]$ [Fig.\,\ref{sort1}-(b)] for all the $k_\text{LCM}(\Phi_m-\Phi_n)$ which falls into a range from 0 to the upper limit of current bin.
In that way, if there is no correlation between $\Phi_m$ and $\Phi_n$,
then we can analytically plot vertical coordinates as a smoothed line, {\it i.e.}, the green dashed line.
We see the red symbols which corresponding to the $\Theta_n'\,(n\leq50)$ randomized case basically obey this rule,
and the little discrepancies are due to the numerical discrete limitation we mentioned in Fig.\,\ref{corrpsi-corrphi-rotation}.
For the events with original participant plane, both $\Phi_2$-$\Phi_3$ and $\Phi_2$-$\Phi_4$ correlations go far away from the green line as the bin number goes up,
which implies that a strong initial participant-plane correlation exists around $|\Phi_m-\Phi_n|=\pi/k_{\text{LCM}}$.
Unlike the initial state participant-plane results, their corresponding final correlations $\langle\cos[k_{\text{LCM}}(\Psi_m-\Psi_n)]\rangle$
show weak dependence with the horizontal bin even if they show large initial correlations.
It further confirms the conclusions revealed by Fig.\,\ref{corrpsi-corrphi-rotation}, {\it i.e.}, the strong final event-plane correlations mainly come from the hydrodynamical transformation of a large eccentricity at the initial state, rather than their strong initial participant-plane correlations.
The rest of Fig.\,\ref{sort1} shows the correlations between $\varepsilon_m$($v_m$) and $\varepsilon_n$($v_n$),
in the way of scattering all the concerned events in $\varepsilon_m$-$\varepsilon_n$(or $v_m$-$v_n$) plane.
Considering the statistics, the scatters are shown in two colors: the light color (magenta and cyan) represents the events which belong to the tenth bin, and the dark color (dark purple and dark blue) represents the events which belong to the other bins.
Making such a classification can make sure that both the two colors have nearly the same number of events, and also we have seen that the tenth bin show strong initial correlations.
The black and red triangles, corresponding to the darker and lighter scatters respectively,
are the average value of $\varepsilon_n$($v_n$) and $\varepsilon_m$($v_m$) in each $\varepsilon_m$($v_m$) bin.
We first analyze the mid-row, which are for the events with original participant plane.
The trends of these data are similar to the ones shown in Ref.\,\cite{NuclPhysA.931.959,PhysRevC.93.014909}.
The $\varepsilon_2$ and $\varepsilon_3$ show a slightly anti-correlation for all these event points,
only that the event with strong initial $\Phi_2$-$\Phi_3$ correlation is also accompanied by a relative large $\varepsilon_3$,
see the red triangles are wholly higher than the black ones.
But from the final state $v_2$-$v_3$ plane, the initial $\Phi_2$-$\Phi_3$ correlation shows having no influence on the relevance between $v_2$ and $v_3$ because that the red triangles are almost overlap with the black ones.
For $\varepsilon_2$ and $\varepsilon_4$, similar to their strong initial $\Phi_2$-$\Phi_4$ correlation which is caused by the frequency doubling effect, they show a strong relevance.
The events which have strong $\Phi_2$-$\Phi_4$ correlations, also have large $\varepsilon_2$ and $\varepsilon_4$.
In $v_2$-$v_4$ panel, strong relevance still survives, but all the red triangles are lower than the black triangles which is in accordance with the experimental results\,
\cite{NuclPhysA.931.959,EurPhysJC.73.2510,PhysRevLett.107.032301,PhysLettB.708.249,EurPhysJC.72.2012,NuclPhysA.904.511c,NuclPhysA.910.276,PhysRevC.92.034903,PhysLettB.773.68}.
In the bottom row, we show the results for $\Theta_n'\,(n\leq50)$ randomized events.
All points in $\varepsilon_2$-$\varepsilon_3$ and $\varepsilon_2$-$\varepsilon_4$ panels are
exact the same ones as the mid-row panels but the colors change to the new classification bins which they belong to.
In the new color representation, the red and black triangles in $\varepsilon_2$-$\varepsilon_3$ panel almost overlap those in $\varepsilon_2$-$\varepsilon_4$ panel as we expected.
All the values of $v_2$, $v_3$ and $v_4$ are smaller than the ones of the events with original participant-plane for nearly 50\%, which reflect that the strong initial participant-plane correlations will transform into large flow coefficients.
However, the trends of the final flow data for the two colored events are similar with the original participant plane events, which means the initial plane correlation almost has no influence on the relevance of the flow coefficients between different orders.

\begin{figure}
\begin{center}
  \includegraphics[scale=0.5]{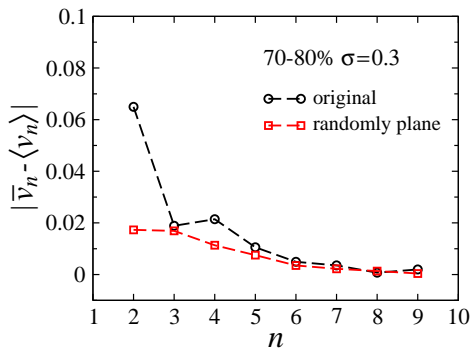}
  \caption{(Color online)
  The difference between $\langle v_n\rangle$ from the event-by-event hydrodynamical calculation
  and $\bar{v}_n$ from the single-shot hydrodynamical calculation for 70-80\% collision centrality range with $\sigma=0.3$ fm.
  Black circle: for the original participant-plane events, red rectangle: for the $\Theta_n'\,(n\leq50)$ randomized events.}
  \label{fourierrot}
\end{center}
\end{figure}

At the end of this section,
we give a brief discussion about the relations between $\langle v_n\rangle$ of the event-by-event hydrodynamical calculation
and $\bar{v}_n$ of the single-shot hydrodynamical calculation for the $\Theta_n'\,(n\leq50)$ randomized events.
In Fig.\,\ref{fourierrot}, we show the relative value of $\langle v_n\rangle$ and $\bar{v}_n$ for two event sets with $\sigma=0.3$ fm.
It shows that the differences between $\langle v_n\rangle$ and $\bar{v}_n$ for the randomized event set are less than the ones for the original participant-plane event set, especially for $v_2$ and $v_4$.
The similarities between $\langle v_n\rangle$ and $\bar{v}_n$, {\it i.e.}, $|\bar{v}_n-\langle v_n\rangle|/\langle v_n\rangle$, can increase nearly 50\% for $v_2$ and $v_4$.
It implies that, in small $\sigma$ peripheral collisions which have large initial eccentricities (see the analysis for Fig.\,\ref{ebe-flow}), the strong initial participant-plane correlations play an important role on increasing the discrepancies of the flow coefficients between the single-shot and the event-by-event simulations.

\section{Summary and Conclusion}

We use Monte Carlo Glauber initial condition and 2+1 viscous hydrodynamics to
investigate the hydrodynamical transformation from the initial geometry representation to the final momentum representation,
for the Pb+Pb $\sqrt{s_\text{NN}}=2.76$ TeV collisions with 0-10\%, 40-50\% and 70-80\% centrality intervals at the LHC.
By constructing a randomized set event on the basis of the original events which generated by the Monte Carlo Glauber code,
we give a detailed cross discussions concerning four quantities, {\it i.e.}, initial state eccentricities, participant-plane correlations, final state flows and participant-plane correlations, with their higher order effects together.

We first examine the final flow coefficients of single-shot and event-by-event simulations by taking four values of the Gaussian width $\sigma$ in each centrality intervals.
We find the equivalence of two simulations gradually breaks in small $\sigma$ peripheral collisions, {\it i.e.}, the cases with large eccentricities.
To understand the event-by-event hydrodynamical transformations, the initial participant-plane and final event-plane correlations in small $\sigma$ peripheral collisions are calculated.
The results reveal a rotation behaviour of each participant-plane during the evolution.

To figure out the transformation effect of the participant-plane correlations,
we construct a new set of events on the basis of the previous original events.
{\it i.e.}, the eccentricities for different orders are kept but their corresponding participant-planes are rotated to new random directions.
Under this operation, the strong participant-plane correlations showed in original event set are eliminated in the new event set, but the final event-plane correlations are still exist, especially for two order numbers in which one is a multiple of the other.
It reveals the strong final event-plane correlations mainly come from the hydrodynamical transformation of the large eccentricity $\varepsilon_n$ at the initial state.
This conclusion is obtained by using the idea of variable-controlling approach.
For example, to figure out the influences of the initial state participant-plane correlations on the final state event-plane correlations, we eliminate the initial state participant-plane correlations while keep their corresponding eccentricities unchanged, {\it i.e.}, randomly rotating each order participant-plane.
In this way, we find that no matter how weak the initial state participant-plane correlations are, the final state event-plane correlations are still strong for the peripheral collisions which carry large initial eccentricities.
Similarly, if we want to figure out the influences of initial eccentricities on the final state correlations, we should eliminate the initial eccentricities while keep their corresponding participant-plane correlations unchanged.
However, this kind of event has no physical meanings. Fortunately, the influences of initial eccentricities can be qualitatively inferred from 0-10\% collisions in Fig.\,\ref{ebe-ecc} and Fig.\,\ref{corrpsi-corrphi}.
It shows that the final state event-plane correlations are weak for the central collisions which carry small eccentricities and weak participant-plane correlations.
Considering that the participant-plane correlations have little effect on the event-plane correlations.
Therefore, we qualitatively conclude that the strong final event-plane correlations mainly come from the hydrodynamical transformation of the large eccentricities at the initial state, rather than their strong initial state correlations.

In addition, considering the complicate relations between the initial eccentricities, the participant-plane correlations, the final flow coefficients and the event-plane correlations, we also give a detailed cross discussions concerning the four quantities in the correlations between the second and third orders, and the ones between the second and fourth orders.
The $v_2$, $v_3$ and $v_4$ of the plane randomized events are smaller than those of the original participant-plane event for nearly 50\%, which reflects that the strong initial participant-plane correlations will transform into large flow coefficients in peripheral collisions.
Comparing with the scattering flow data for the plane randomized events and the original participant-plane events, we also find that the initial state participant-plane correlation almost has no influence on the relevance of the flow coefficients between two different orders.

Our results and discussions imply that, in small $\sigma$ peripheral collisions which have large initial eccentricities, the strong initial participant-plane correlations play an important role on increasing the discrepancies of the flow coefficients between the single-shot and the event-by-event simulations.

In this paper, we only focus on one kind model of initial conditions, {\it i.e.}, MC-Glauber model.
One may concerns the sensitivity of our results to model uncertainties in the initial state.
For other initial condition models with the same calculation process, their results will show great similarities in qualitative interpreting the hydrodynamical transformation effect from the initial participant-plane correlations, and also in the relations between single-shot and event-by-event hydrodynamical calculations.
Besides, the early pre-equilibrium effects of the system are not considered here.
The early pre-equilibrium expansions will smear out the early stage fluctuations and generate some amount of early flow\,\cite{PhysRevC.82.064903}, it may affect our results especially for peripheral small $\sigma$ events.
In future, a quantitatively study for pre-equilibrium effect is needed.
In addition, as a branch of machine learning, deep-learning has been recently proved efficient in unveiling hidden information from the highly implicit data of heavy-ion experiments\,\cite{NuclPhysA.982.927,NatureComm.9.210}.
It shows great potential in mimicking the evolutions of hydrodynamics, but directly extracting the initial information from the final experimental data is still a great challenge.
Our study could provide additional hydrodynamical informations for the training process of deep-learning, which will be helpful for mimicking the hydrodynamical transformations from the initial state to the final state.
Therefore, our results has potential to be applied into the deep-learning study in the future.

\begin{acknowledgments}
This work is supported by National Natural Science Foundation of China (11747155 and 11647166), Science and Technology Foundation of Education Department of Jilin Province of China (JJKH20181162KJ), Natural Science Foundation of Inner Mongolia (2017BS0104).
\end{acknowledgments}


\end{document}